\documentclass[preprint]{aastex}

\def\mic{{$\mu$m}}

\def\h2o{H$_2$O}

\def\aple{$\mathrel{\hbox{\rlap{\hbox{\lower4pt\hbox{$\sim$}}}\hbox{$<$}}}$}

\def\apge{$\mathrel{\hbox{\rlap{\hbox{\lower4pt\hbox{$\sim$}}}\hbox{$>$}}}$}

\begin{document}

\title{2 \mic \ Narrow--band Adaptive Optics Imaging in the Arches Cluster}

\author{R. D. Blum\altaffilmark{1}} \affil{Cerro Tololo Interamerican
Observatory, Casilla 603, La Serena, Chile\\ rblum@noao.edu}

\author{D. Schaerer\altaffilmark{1}} \affil{ Laboratoire
d'Astrophysique, Observatoire Midi-Pyr\'en\'ees, 14 Av. E. Belin F-31400,
Toulouse, France \\schaerer@ast.obs-mip.fr}

\author{A. Pasquali} \affil{ESO/Space Telescope European Coordinating
Facility, Karl-Schwarzschild-Strasse 2, D-85748 Garching bei München,
Germany \\ apasqual@eso.org}

\author{M. Heydari-Malayeri} \affil{DEMIRM, Observatoire de Paris, 61
Avenue de l'Observatoire, F-75014 Paris, France\\ heydari@obspm.fr}

\author{P. S. Conti} \affil{JILA, University of Colorado\\Campus Box
440, Boulder, CO, 80309, USA \\pconti@jila.colorado.edu}

\author{W. Schmutz} \affil{Physikalisch-Meteorologisches
Observatorium, 7260 Davos, Switzerland\\ wschmutz@pmodwrc.ch}

\altaffiltext{1}{Visiting Astronomer, Canada--France--Hawaii--Telescope,
operated by the National Research Council of Canada, the Centre National 
de la Recherche Scientifique de France, and the University of Hawaii}

\begin{abstract}

Canada--France--Hawaii-Telescope adaptive optics bonnette images
through narrow--band filters in the $K-$band are presented for the
Arches cluster. Continuum fluxes, line fluxes, and equivalent widths
are derived from high angular resolution images, some near diffraction
limited, for the well known massive stars in the Arches
cluster. Images were obtained in the lines of \ion{He}{1} 2.06 \mic,
\ion{H}{1} Br$\gamma$ (2.17 \mic), and \ion{He}{2} 2.19 \mic \ as well
as continuum positions at 2.03 \mic, 2.14 \mic, and 2.26 \mic. In
addition, fluxes are presented for \ion{H}{1} P$\alpha$ (1.87 \mic)
and a nearby continuum position (1.90 \mic) from Hubble Space
Telescope archival data\footnote{"Based on observations made with the
NASA/ESA Hubble Space Telescope, obtained from the data archive at the
Space Telescope Science Institute. STScI is operated by the
Association of Universities for Research in Astronomy, Inc. under NASA
contract NAS 5-26555."}. The 2 \mic \ and P$\alpha$ data reveal two
new emission--line stars and three fainter candidate emission--line
objects.  Indications for a spectral change of one object between
earlier observations in 1992/1993 and our data from 1999 are found.
The ratio of \ion{He}{2} 2.19 \mic \ to Br$\gamma$ emission exhibits a
narrow distribution among the stars, suggesting a narrow evolutionary
spread centered predominantly on spectral types O4~If or Wolf-Rayet
stars of the WN7 sub--type.  From the approximate spectral types of
the identified emission--line stars and comparisons with evolutionary
models we infer a cluster age between $\sim$ 2 and 4.5 Myr.
\end{abstract}

\keywords{instrumentation: adaptive optics --- Galaxy: center ---
stars: early--type --- stars: Wolf--Rayet}

\section{INTRODUCTION}

The Arches cluster is thought to be the most massive, dense, star
forming cluster in the Galaxy \citep{ssf98, feal99}. This prolific
starbursting cluster which may harbor more than one hundred O--type
stars \citep{ssf98} and whose linear size at 8 kpc is only about 0.6
pc in diameter \citep{feal99} was first found in limited near infrared
surveys by \citet{neal95} and \citet{ceal96}. \citet{neal95} named it
``Object 17,'' for such it was in their spectroscopic survey near the
Galactic center. \citet{ceal96} referred to the cluster as G0.12+0.2
after its position on the sky, but the cluster has become known as the
``Arches'' taking the name of the Galactic center thermal Arched
Filaments, thermal emission structures 30 pc in projection from the
Galactic center \citep{myz85}. The Arches cluster is located at $\alpha (2000)
= 17^{\rm h} 45^{\rm m} 50.^{\rm s}4 \ \delta (2000) = -28^{\circ} 49'
22''$ \citep{feal99}.

\citet{neal95} presented modest angular resolution images of the
Arches cluster, and \citet{ceal96} presented low resolution $K-$band spectra
for 13 of the Arches cluster's brightest stars. Both investigations showed the
Arches cluster contained massive emission--line stars similar to those which
had been discovered previously in the Galactic center
\citep{feal87,ahh90,keal91}. \citet{feal99} used Hubble Space
Telescope (HST)/NICMOS images to explore the mass function in the
Arches cluster. They found that the Arches cluster IMF is weighted much more to
massive stars than ``typical'' mass functions, e.g. \citet{s55}. It
has been argued that massive stars might preferentially form near the
center of the Galaxy due to the pre-conditions there \citep{ms96}. The
mass function in the Arches cluster thus stands in stark contrast to the OB
associations in the Galaxy and Large Magellanic Cloud
\citep[LMC]{mh98}, including the massive star cluster R136 in the LMC.

Though there is ample evidence that the Arches cluster harbors a dense cluster
of massive stars, and the brightest are apparently evolved as
evidenced by their their emission--line spectra \citep{ceal96}, the
precise mass scale has yet to be set by detailed models of the stars
or comparison to evolutionary tracks of stellar parameters determined
from spectroscopic measurements. In this paper, we present high
angular resolution 2 \mic \ narrow--band images (which cover 36$''$ in
a single frame) taken at the Canada--France--Hawaii-Telescope (CFHT)
and 1.9 \mic \ (archival) images taken with the Hubble Space Telescope
(HST)/NICMOS. This is the first of two papers in which we will explore
the massive stars in the Arches cluster. Here we present our data and
observational measurements. In a subsequent paper we will use the
accurate line and continuum fluxes determined from these high angular
resolution observations to interpret stellar models and further
quantify the properties of the Arches cluster emission--line stars.

\section{OBSERVATIONS AND DATA REDUCTION}

\subsection{CFHT Adaptive Optics Images}

The Canada--France--Hawaii-Telescope (CFHT) images were obtained on
the nights of 29 and 30 June 1999 and 01 July 1999 using the facility
infrared imager, KIR, and the adaptive optics bonnette (AOB),
PUEO. KIR employs a 1024$\times$1024 HgCdTe HAWAII detector with
0.035$''$ pixels (36$''$ FOV). KIR and PUEO are fully described by
\citet{bh98}; see also \citet{real98}. The Arches cluster data were
obtained on 01 July, while observations of Wolf--Rayet (WR) stars and
photometric standards were obtained on all three nights. All data
obtained on the Arches, WR stars, and standards were obtained with the
adaptive optics (AO) corrections on.

Emission--line images were taken through narrow--band filters centered
at 2.183 \mic \ (\ion{He}{2}, a red-shifted Br$\gamma$ filter,
$\Delta\lambda \approx 300$ \AA), 2.17 \mic \ (Br$\gamma$,
$\Delta\lambda \approx 200$ \AA), 2.06 \mic \ (\ion{He}{1},
$\Delta\lambda \approx 280$ \AA). Continuum images were obtained at
2.26 \mic \ ($\Delta\lambda \approx 600$ \AA), 2.14 \mic \
($\Delta\lambda \approx 300$ \AA), and 2.03 \mic \ ($\Delta\lambda
\approx 210$ \AA).

\subsubsection{Basic Reduction and Standard Stars}

All basic data reduction was accomplished using IRAF\footnote{IRAF is
distributed by the National Optical Astronomy Observatories.}. Each
image was flat--fielded using dome flats and then sky subtracted using
a median combined image of five to six frames. For the Arches itself,
independent sky frames were obtained $\sim$ 19 arcminutes south--east
of the cluster. Wolf--Rayet and standard stars used the median
combination of the data for sky.

Each night three standard stars were observed. Two of the standards
were chosen to span a large range of airmass (approximately one to two
airmasses) and observed sequentially over a short time interval. The
set of three standards also spanned the observing period of the Arches
cluster. The Arches data and Wolf--Rayet stars have been placed on a
flux basis by interpolating over the standard star observations as a
function of airmass.

The standard stars observed are all early A--type stars: HD130163
(A0V), HD201941 (A2), and HD205772 (A5IV/V). The standard star
narrow--band magnitudes were placed on a flux scale by adopting the
2.2 \mic \ flux corresponding to the $K-$band magnitude \citep{efmn82}
and assuming the relative flux at the filter effective wavelength is
given by a 9800 K black body \cite{j66}. The flux at 2.2 \mic \ is for
$\alpha$ Lyra, 4.07$\times$10$^{-10}$ W--m$^{-2}$--\mic$^{-1}$
\citet{at86}. On any given night, the difference between the standard
star instrumental magnitudes and the assigned flux was less than 5$\%$
for all stars observed at similar airmass in all filters.  Typically,
the difference was about 3$\%$ rms. The derived airmass corrections
varied between about 0.01 and 0.10 magnitudes per airmass depending on
the filter.  The 2.03 \mic \ and 2.06 \mic \ filters showed the
largest corrections (0.07 mag and 0.10 mag per airmass) as is expected
since this is where the opacity in the Earth's atmosphere is strongly
affected by H$_2$O and CO$_2$ molecules.

The resulting point spread functions (PSF) on the final combined
CFHT images
exhibit full widths at half maximum (FWHM) intensity between 4.9 (the 
2.19 \mic \ image)
pixels to 8.0 pixels (the 2.26 mic \ image). 
The typical image quality for any of the six final images
was thus 0.17$''$ to 0.28$''$ FWHM. The best image quality
was obtained on the 2.19 \mic \ image where the airy diffraction
pattern can bee seen. The minimum airmass for any of the Arches images
was 1.5 and some of the images were obtained at 2.0 airmasses, the AOB
delivering excellent correction even at such high zenith
distance. Several of the standard stars observed near zenith were very
nearly diffraction limited. For example observations of WR138 on 30
June were 3.5 pixels (0.123$''$, $\lambda/D = 0.125''$) FWHM at
in all six filters. Typical
estimates of the seeing at $V$ during our run were near 0.8$''$ with
occasional periods as good as 0.4$''$ or as bad as 1.1$''$. However,
the 2 \mic \ images for bright standards or WR stars typically showed
FWHMs near the diffraction limit even at airmass up to 2.0 if the
seeing at $V$ was better than $\sim$ 1$''$.

\subsubsection{DAOPHOT Analysis}

The instrumental magnitudes for the Arches data were obtained using
DAOPHOT \cite{s87}. The version of DAOPHOT employed is included as
part of the XVISTA package V6.0 \citep{h01}. The data were analized
using average PSFs made from four to five stars in the least crowded
regions from around the frame. The Arches and WR star images were
calibrated using aperture corrections derived from the standard star
data and apertures of between 100 and 130 pixels in radius (7.0$''$ to
9.1$''$ diameter).  Such large apertures were required to obtain all
the flux in the AO corrected images. Aperture measurements were made
on the Arches data for the PSF stars after having subtracted the
remainder of the stars found by DAOPHOT. Since the brighter PSF stars
were located outside the dense core of the cluster, this procedure
results in aperture corrections with uncertainties less than three
percent in the mean. The total uncertainties in the Arches
narrow--band magnitudes are the sum in quadrature of the PSF fitting
errors and the uncertainty in the aperture corrections used to put the
instrumental magnitudes on a flux scale. The aperture correction
uncertainty dominates for the brighter stars and is $\pm$0.014, 0.016,
0.012, 0.025, 0.025, and 0.019 for the 2.26 \mic, 2.19 \mic, 2.17
\mic, 2.14 \mic, 2.06 \mic, and 2.03 \mic \ filters, respectively.

The PSF fitting photometry was conservatively limited by considering
only stars with uncertainties of 0.2 mag or less since the main goal
of this work is to study the brightest emission--line stars at high
angular resolution. This resulted in some \apge 800 stars in each
filter which were subsequently used to construct narrow--band
indices (each index is contructed from three filters: one line and two
continuum). Further constraining each narrow--band index to have an
uncertainty less than $\pm$ 0.2 mag and that stars must match to within one
pixel from filter to filter, resulted in approximately 550 stars for
each narrow--band filter (and thus, in each narrow--band index 
which requires three magnitudes to derive it; see
below) The faintest stars for which narrow--band indices are presented
reach a 2.2 \mic \ continuum magnitude of approximately 17. The level
of completeness is much brighter, especially in the cluster center. 

\subsubsection{Comparison of WR Star Line Strengths to Spectroscopy}

In order to assess the narrow--band filter behavior, WR stars with
strong emission lines in \ion{He}{1}, Br$\gamma$, and \ion{He}{2} were
also observed and line strengths derived compared to literature values
based on spectroscopic measurements. The results are show in
Table~\ref{wr}. Values for equivalent widths ($W_{\lambda}$) from
spectroscopic measurements are from \citet{bds95}, \citet{cs96}, and
\citet{fmn97}.  The $W_{\lambda}$ (
line flux/continuum flux density) for the narrow--band measurements is
derived directly from the narrow--band index which is defined as
index$=$ $(f_{\rm line} - f_{\rm cont})/f_{\rm cont}$ where $f_{\rm line}$ is
the flux density in the narrow--band passband at the line position,
and $f_{\rm cont}$ is the continuum flux density interpolated from the
blue and red continuum points nearest to the line. The $W_{\lambda}$
follows by multiplication by the appropriate emission--line filter
band width, here taken to be the quoted value above for each filter.
The index as defined here is a fraction of the continuum.

For the \ion{He}{1} line, the stars WR122, and WR123 were
observed. WR122 has a very strong \ion{He}{1} 2.06 \mic\ line 
and this was found to
contaminate the 2.03 \mic \ filter (the CFHT 2.03 \mic \ filter curve
suggests the response is a few $\%$ at 2.06 \mic), 
resulting in a continuum which was
too high, hence a line strength which was too low (424 \AA\ compared to
an average value in the literature of 680 \AA; see
Table~\ref{wr}). WR122 is reddened so its continuum is nearly flat at
2 \mic \ \citep{fmn97}. Using only the 2.14 \mic \ line as a
continuum (which is further separated from the \ion{He}{1} line than the
2.03 \mic \ filter), 
the line strength of WR122 is in good agreement with the
literature values (623 \AA compared to 680 \AA). For WR123, the 2.03
\mic \ continuum point is affected by \ion{He}{2} emission, and the
\ion{He}{1} 2.06 \mic \ line itself has both an emission and
absorption (P Cygni) component \citep{fmn97}. The line strength is thus again
underestimated. Using only the 2.14 \mic \ continuum point, the
narrow--band line strength is within $\sim$ 40$\%$ of the published
values, which appears consistent with the observed emission plus
absorption components. These potential problems with the \ion{He}{1} 2.06 \mic \
line are not important in the present case as we find no evidence for
any strong lined \ion{He}{1} emission--line stars in the Arches
cluster.

The line strength results are more straightforward for Br$\gamma$, and
as Table~\ref{wr} indicates in good agreement with the literature. A
Br$\gamma$ measurement was obtained for each of WR122, WR123, WR128, WR131,
WR134, and WR138. The mean difference (as a 
percentage of the literature value) is $+0.07\% \pm 0.08\%$. While
consistent with no difference, we have made an 0.07 magnitude
adjustment to the calibration for Br$\gamma$ since this value is
consistent with the expectation that we have under corrected the
standards due to intrinsic Br$\gamma$ absorption and so over corrected
the WR stars. The sense and magnitude of this correction is in good
agreement with that found (by different means) by \citet{bd99} for
similar narrow--band observations.

The measurements for \ion{He}{2} are also in good agreement with the
literature, with the exception of WR131 (no measurement was made for
WR122). Our narrow--band $W_{\lambda}$ for WR131 is an order of
magnitude brighter than the spectroscopic value given by
\citet{fmn97}. There is no obvious reason why such a strong value
should be measured, except for the possibility that the line has
indeed varied. 
\citet{vdh01} 
have noted that WR131 has "diluted emission lines", which we speculate
may be due to the presence of a companion.
If the dilution is due to continuum emission from the companion,
then the emission lines might vary during an eclipse, but
{\it both} the Br$\gamma$ and \ion{He}{2} lines should vary if this were
the case. In anycase, WR131 is a good candidate for future synoptic 
observations.

\subsection{HST NICMOS Images}

The P$\alpha$ (F187N, $\Delta\lambda \sim 190$ \AA) and 1.90 (F190N,
$\Delta\lambda \sim 190$ \AA) \mic \ continuum images were obtained
from the HST archive. These images were processed by the standard
calnica and calnicb pipelines from the Space Telescope Science
Institute (STScI). The images have a total of 256 seconds exposure
time each, and the platescale is 0.076$''$ pix$^{-1}$. Broad--band
images from the same HST program were presented for the Arches by
\citet{feal99}.

Instrumental magnitudes were obtained by DAOPHOT in the same way as
for the CFHT data described above. The photometry was calibrated using
aperture corrections derived from six pixel radius apertures (0.9$''$
diameter) and the flux calibration of \citet{mr01}. The precise values
used were 4.107$\times$10$^{-5}$ and 4.455$\times$10$^{-5}$
Jy--ADU$^{-1}$--s$^{-1}$, respectively for the F187N and F190 filters.
Zero points for the two filters are 825 and 808 Jy \citep{mr01}. The
HST and CFHT magnitudes are not on the same photometric system, but
the line and continuum fluxes are, of course, directly comparable.

The NICMOS data were limited by photometric error in the same way as
described above for the CFHT data. This resulted in about 220 stars in
each filter and some 200 stars with a P$\alpha$ emission--line
index. In principle, the narrow--band index results in an
$W_{\lambda}$ as determined above for the CFHT data, but here the
continuum value is taken directly from the 1.90 \mic \ filter. The
faintest stars for which a P$\alpha$ index is presented reach about
16 mag.

\section{RESULTS and DISCUSSION}

Representative images in the lines of \ion{He}{2} (2.19 \mic) and
P$\alpha$ \ are shown in Figure~\ref{a218} and \ref{f187},
respectively. The other line and continuum images are not shown
(except 2.26 \mic; see below) but are similar.  Positions (offsets
from the AO guide star; see Figure~\ref{a218}) for the brightest stars
with narrow--band magnitudes in all eight filters (hence four emission
line measurements with one exception discussed below) are given in
Table~\ref{pos}. We have arbitrarily cut this list at a 2.26 \mic \
magnitude limit of approximately 12 magnitudes (see
Table~\ref{mags}). The complete photometry set is available upon
request from the authors.

\subsection{Narrow--band Indices}

In Figures~\ref{nb218}, \ref{nb216}, \ref{nb206}, and \ref{nb187} the
narrow--band index is plotted for each emission line. Inspection of the plots
suggests the absolute and relative fluxes are quite good: typical
stars have very nearly zero line flux (a zero index). The rms
deviation (excluding 3 sigma outliers to a linear fit) is 5, 6, 5, and
3 $\%$ for the \ion{He}{2} 2.19 \mic, Br$\gamma$ 
2.17 \mic, \ion{He}{1} 2.06 \mic, and P$\alpha$ 1.87 \mic \
indices, respectively. For the CFHT data, crowding clearly begins to
dominate the indices (i.e. the photometry) between 14th and 15th
magnitudes.

There are small, systematic, offsets from zero evident in the plots.
Such offsets could be due to the intrinsic absorption in the
atmospheres of the Arches cluster hot stars, or due to unaccounted for
offsets in one or more filters. As described in the following, the
most likely case is the latter; however, the knowledge of the expected
strength of these features in O and B stars \citep{hcr96} can be used
to correct the mean indices.  The mean index between 12th and 14th
magnitude for the \ion{He}{2} 2.19 \mic, Br$\gamma$ 2.17 \mic, and
\ion{He}{1} 2.06 \mic \ indices is $-$3, $-$6, and $-$2 $\%$,
respectively. The P$\alpha$ 1.87 \mic \ index is approximately zero
near 12.5 magnitudes and falls to a minimum of approximately $-$12$\%$
at fainter magnitudes. The P$\alpha$ line index is discussed below;
here we concentrate on the 2 \mic \ indices. Main sequence O--type
stars exhibit absorption in each of the 2 \mic \ features
\citep{hcr96} discussed here, so negative indices might be
expected. At 14th magnitude, these objects should be mid to late O, or
perhaps early B--type stars \citep{ssf98}. The absorption strengths
for such objects are relatively constant for all three features in
this spectral type range \citep{hcr96}, so we can use the known
properties of such stars to verify or set the mean level of the index
in each feature for the Arches cluster stars. From the data presented
by \citet{hcr96}, an estimate of the mean absorption is 1, 4, and 2
\AA $W_{\lambda}$ for the \ion{He}{2} 2.19 \mic, Br$\gamma$, and
\ion{He}{1} 2.06 \mic \ lines, respectively. This translates to index
strengths of 0.003, 0.02, and 0.007. This is less absorption than the
mean indices imply. Therefore, the indices for each of the 2 \mic \
features have been corrected {\it to} these mean values; i.e. the
amount of {\it absorption} in the \ion{He}{2} 2.19 \mic, Br$\gamma$
2.17 \mic, and \ion{He}{1} 2.06 \mic \ indices has been {\it reduced}
by a constant amount equal to 3, 4, and 2 $\%$ respectively in the
final determination of the line fluxes shown in Table~\ref{flux} (the
indices in Figures~\ref{nb218}, \ref{nb216}, and \ref{nb206} are the
uncorrected, values).

The offset for the P$\alpha$ index is somewhat larger, $-12\%$. By
analogy to the Br$\gamma$ index, systematic absorption at P$\alpha$ 1.87 \mic \
is expected. However, the situation is not as straight forward since
there is no blueward continuum point: part of the systematic negative
offset seen in Figure~\ref{nb187} might be due to reddening. For
heavily reddened objects toward the Galactic center, the continuum
actually increases to the red, even for these hot stars. The ratio of
blackbody fluxes on the Rayleigh--Jeans tail at 1.87 and 1.90 \mic \
results in an index of $~\sim$ $-$5$\%$ for an interstellar extinction
of $A_K =$ 3.1 mag appropriate to the Arches \citep{feal99} and using
a typical interstellar extinction curve \citep{m90}.  Furthermore, to
our knowledge, no study of the intrinsic P$\alpha$ absorption in O and
B stars exists. Due to the uncertain contribution to the index from
these systematic effects, no correction similar to those for the 2
\mic \ indices has been made to the P$\alpha$ index.

\subsection{The Emission--line Stars}

\begin{figure}
\figurenum{6} 
\plotone{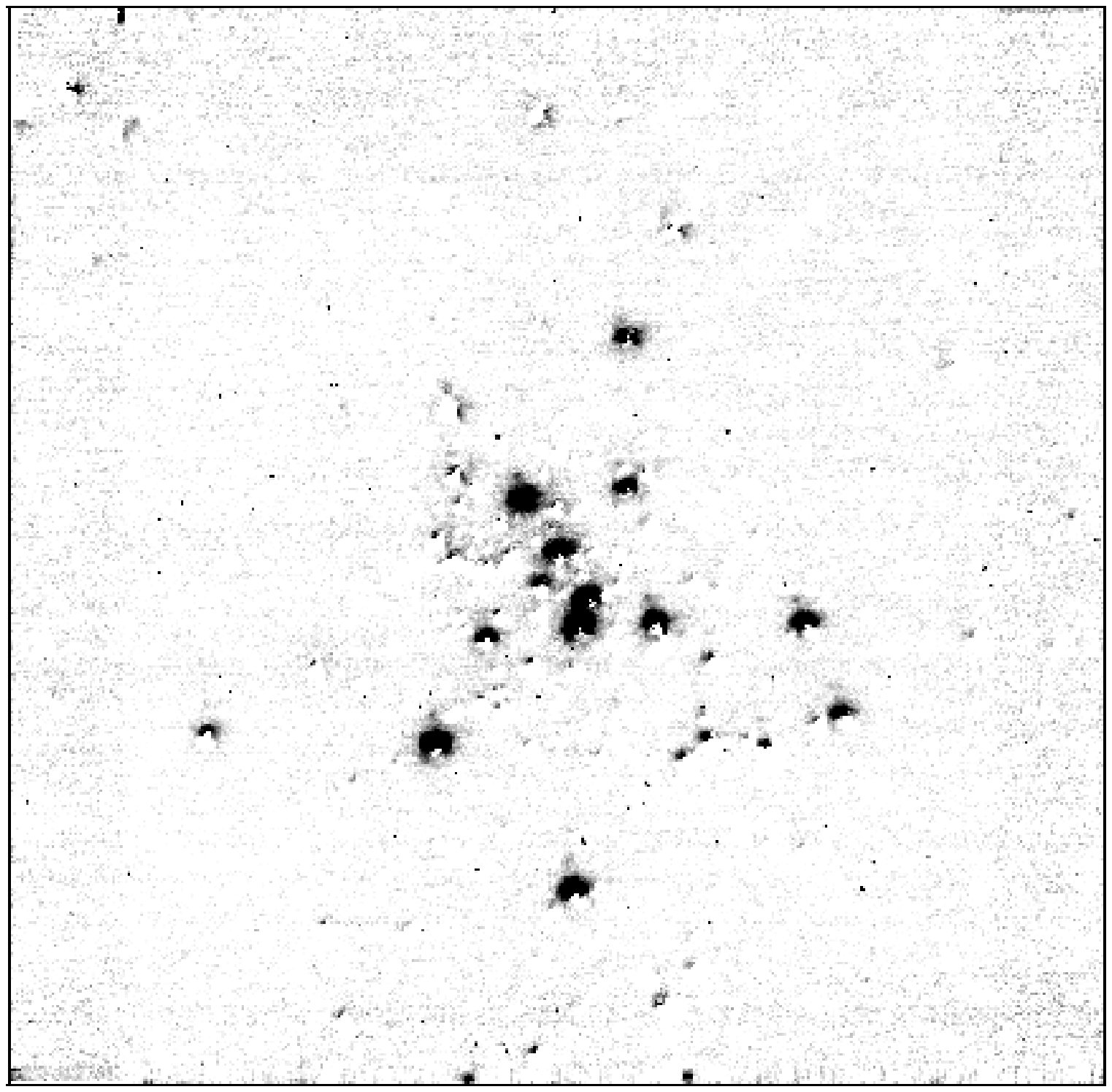}
\figcaption[]{Br$\gamma$. Continuum subtracted 2.17 \mic \ image using
the 2.26 \mic \ continuum (for simplicity). The adaptive optics guide
star appears as a false emission--line star due to its strong blue
continuum, but the detailed photometry using two continuum points
accounts for this; see text. The bright object to the south of the
image is a newly identified emission--line star.}
\label{brg}
\end{figure}

\begin{figure}
\figurenum{7} 
\plotone{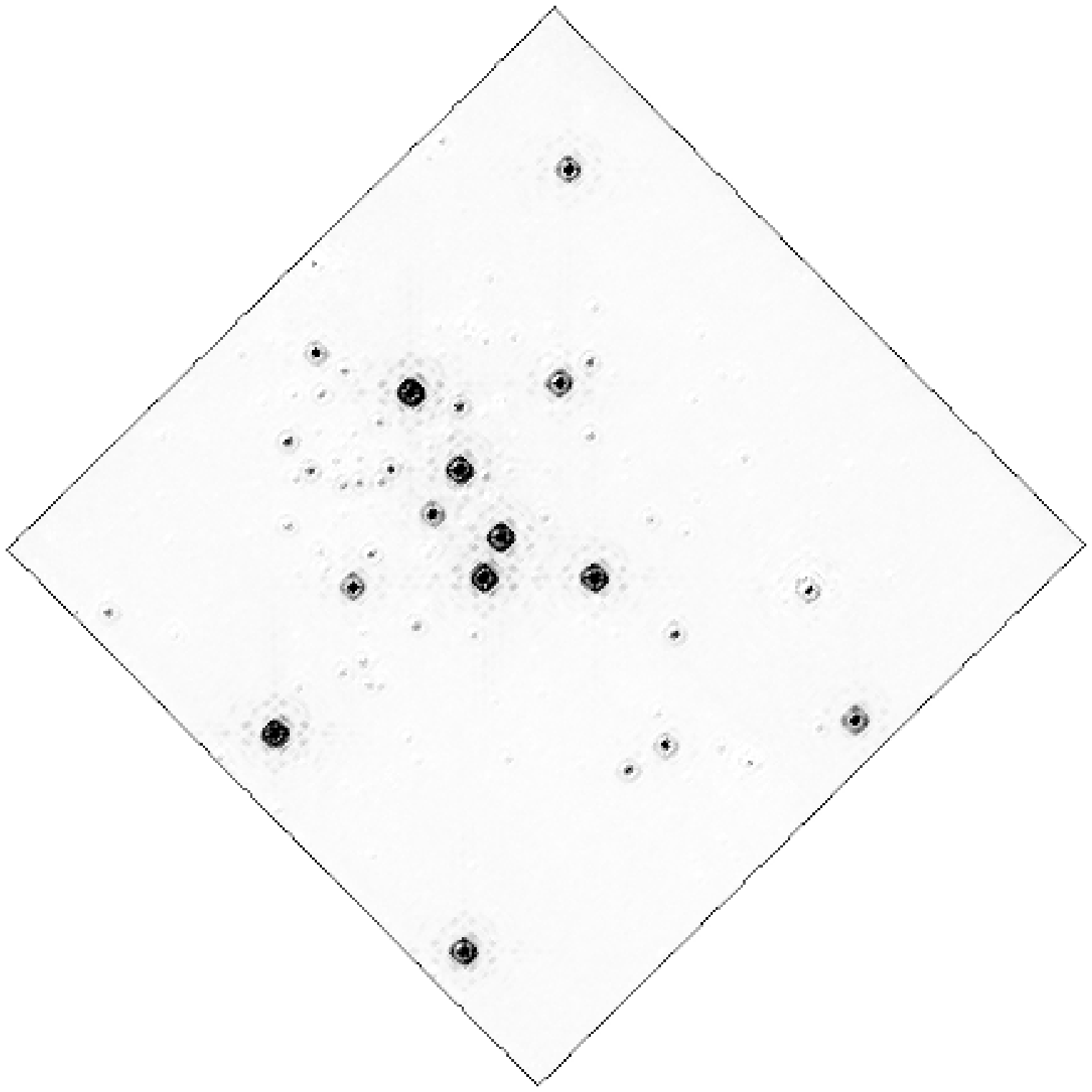} 

\figcaption[]{P$\alpha$ 1.87 \mic \ continuum subtracted image. The
bright object near the bottom is a newly identified emission--line
star.}

\label{paa}
\end{figure}

The present data set was obtained chiefly to provide better spatial
resolution, and hence more accurate line and continuum measurements,
for the bright emission--line stars previously discovered in the
Arches cluster by \citet{ceal96}. A graphical demonstration of the
improvement obtained by AO and/or HST imaging is shown in
Figures~\ref{brg} and \ref{paa}. Each individual emission--line object
is well separated from the others as (well as with the other
non-emission--line sources). This was not the case in images presented
by \citet{neal95} and \citet{ceal96}.  These are continuum subtracted
line images. For simplicity, the Br$\gamma$ image is constructed
using only the red continuum. Both of these images show that continuum
variations can produce spurious emission--line objects; the AO guide
star appears due to its strong blue continuum. Our technique of using
two continuum filters avoids this problem in general. The AO guide
star is also obviously a foreground object based on color and a
$K-$band spectrum \citep{ceal96}. The HST PSF and pointing are
remarkably stable. The P$\alpha$ image does not show the effects of
seeing and scale variations as the ground--based image does.

The set of emission--line stars identified by \citet{neal95} and
\citet{ceal96} can be easily recognized in Figures~\ref{brg} and
\ref{paa}. The Cotera and Nagata IDs are given in
Table~\ref{flux}. The positions of all the objects presented in
Table~\ref{flux} are marked on the 2.26 \mic \ continuum image
(Figure~\ref{a226}).

Comparing the present line fluxes to \citet{ceal96}, we find a
difference of $-30 \pm 70\%$ in the Br$\gamma$ line.  For the stars
most isolated (Cotera ID: 1, 9, 10, 11, 13) the fluxes presented by
\citet{ceal96} are in better agreement with those determined here
(Table~\ref{flux}). The RMS difference is $11 \pm 33\%$. Much of this
difference is due to Cotera star 11 which is different by a factor of
two (more flux in the present data). Considering stars 1, 9, 10, and
13, the fluxes agree to $0 \pm 26\%$. \citet{neal95} present
Br$\gamma$ and Br$\alpha$ fluxes for eight of the 13 Cotera
stars. Their lower angular resolution measurements are also generally
in good agreement with our Br$\gamma$ fluxes. We find a difference of
$-40 \pm 50\%$, which reduces to $ -20 \pm 20\%$ if we disregard
Cotera star 13 in the comparison which is different to our measurement
by a factor of 2.

For the remainder of the \citet{ceal96} Br$\gamma$ measurements, there
are three objects which differ in flux by a factor of two to three, in
each case the Cotera flux being larger, and these are three of the
four stars with the largest fluxes reported by \citet{ceal96}, Cotera
stars 2, 3, and 6. The \citet{neal95} values for these objects are
actually more consistent with our measurements. The \citet{ceal96}
spectra were extracted using relatively large sky offsets. It is
possible that there is nebular Br$\gamma$ emission contaminating their
spectra. The sky values for each of the filters presented here were
determined ``locally'' as a free parameter in the DAOPHOT PSF fits. It
is not clear whether or not local sky values were determined by
\citet{neal95}.

\begin{figure}
\figurenum{2} 
\plotone{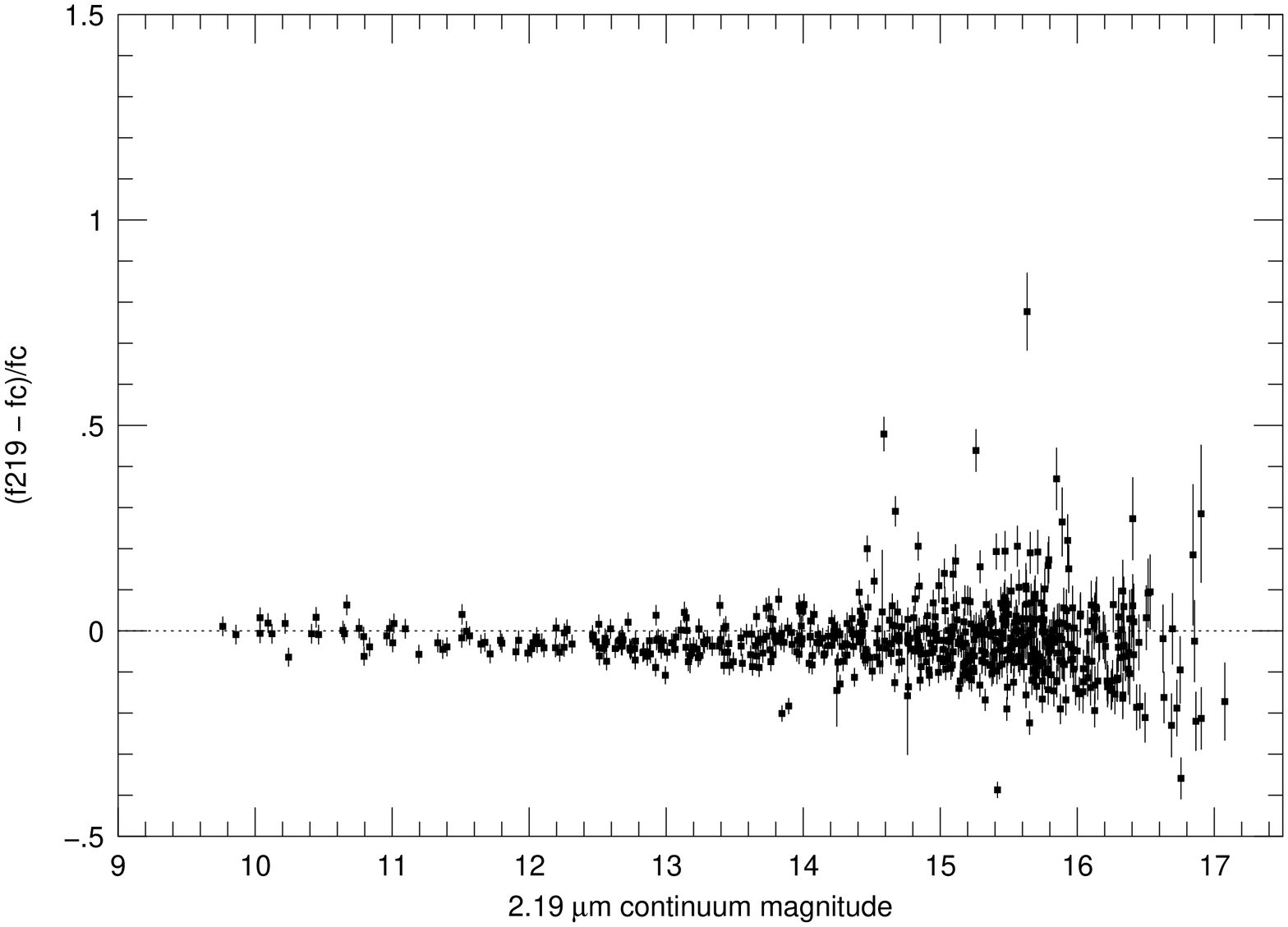} 
\figcaption[]{\ion{He}{2} 2.19 \mic \ Narrow--band index. The index is similar
to an equivalent width; see text. Line fluxes and equivalent widths in 
Tables~\ref{flux} and \ref{equiv} have been corrected for the systematic 
``absorption'' evident in this plot; see text.}
\label{nb218}
\end{figure}

\begin{figure}
\figurenum{3} 
\plotone{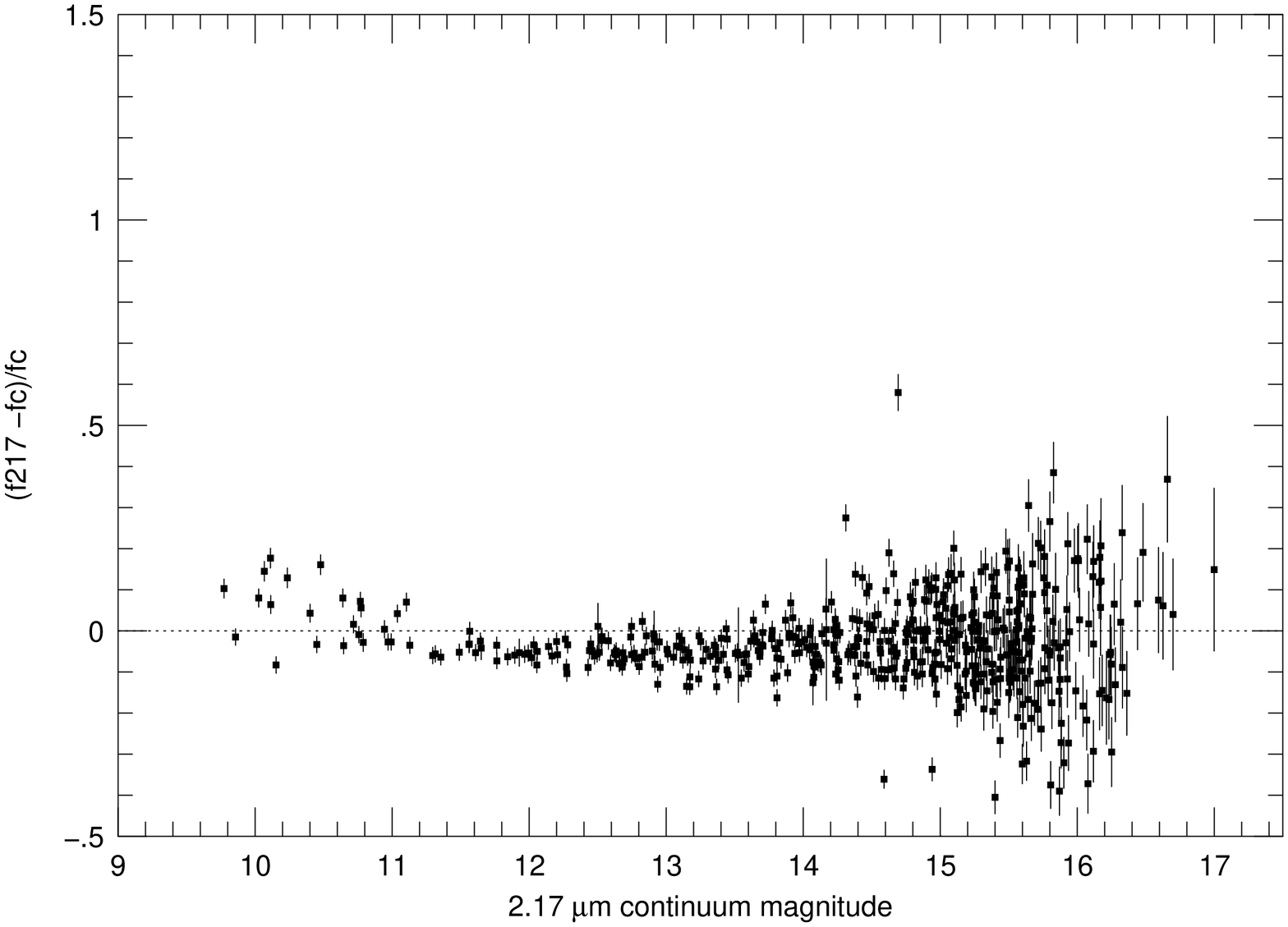} 
\figcaption[]{Same as Figure~\ref{nb218} but for the Br$\gamma$ line.}
\label{nb216}
\end{figure}

\begin{figure}
\figurenum{4} 
\plotone{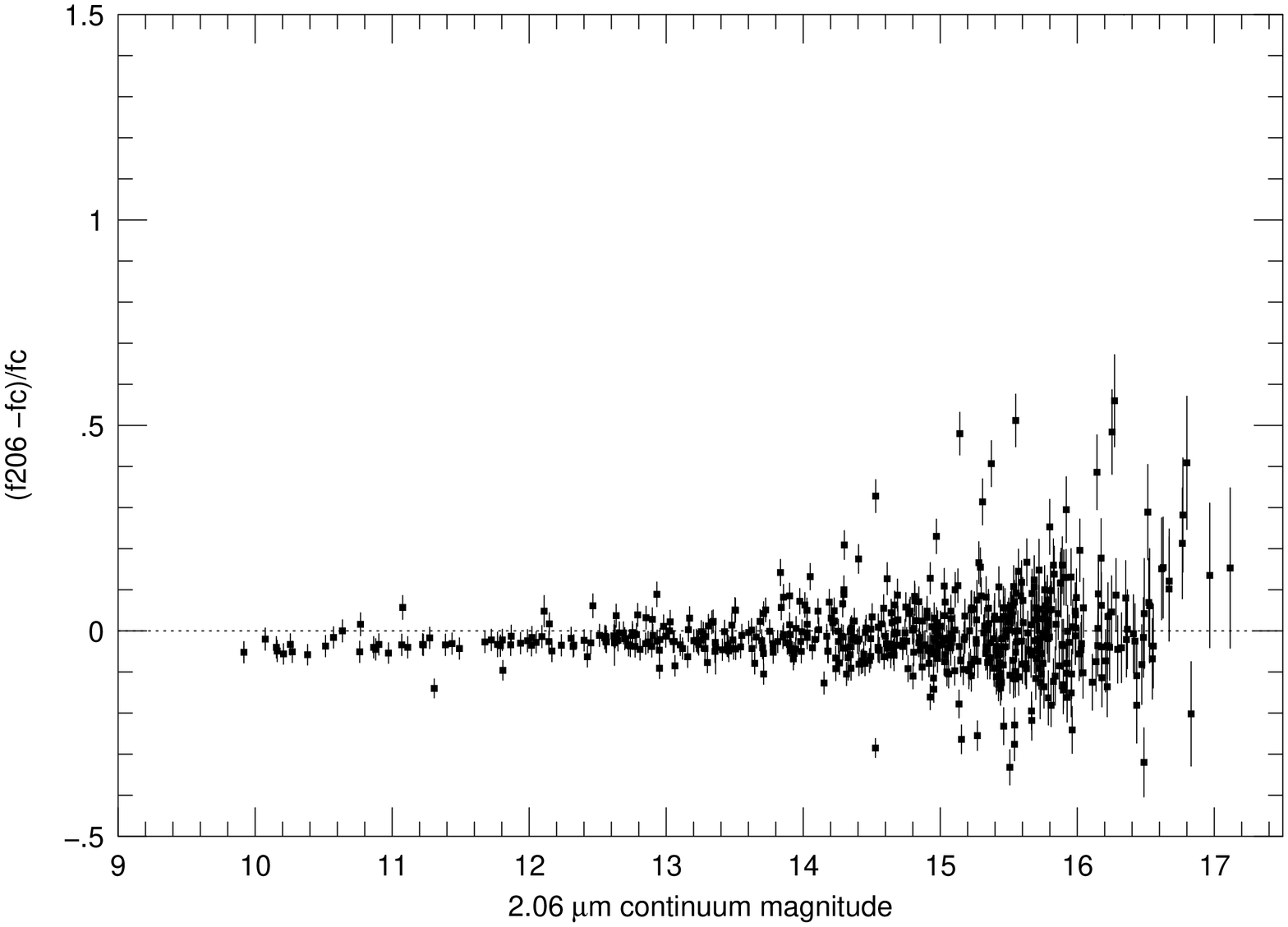} 
\figcaption[]{Same as Figure~\ref{nb218} but for the \ion{He}{1} 2.06 \mic \ 
line.}
\label{nb206}
\end{figure}

\begin{figure}
\figurenum{5} 
\plotone{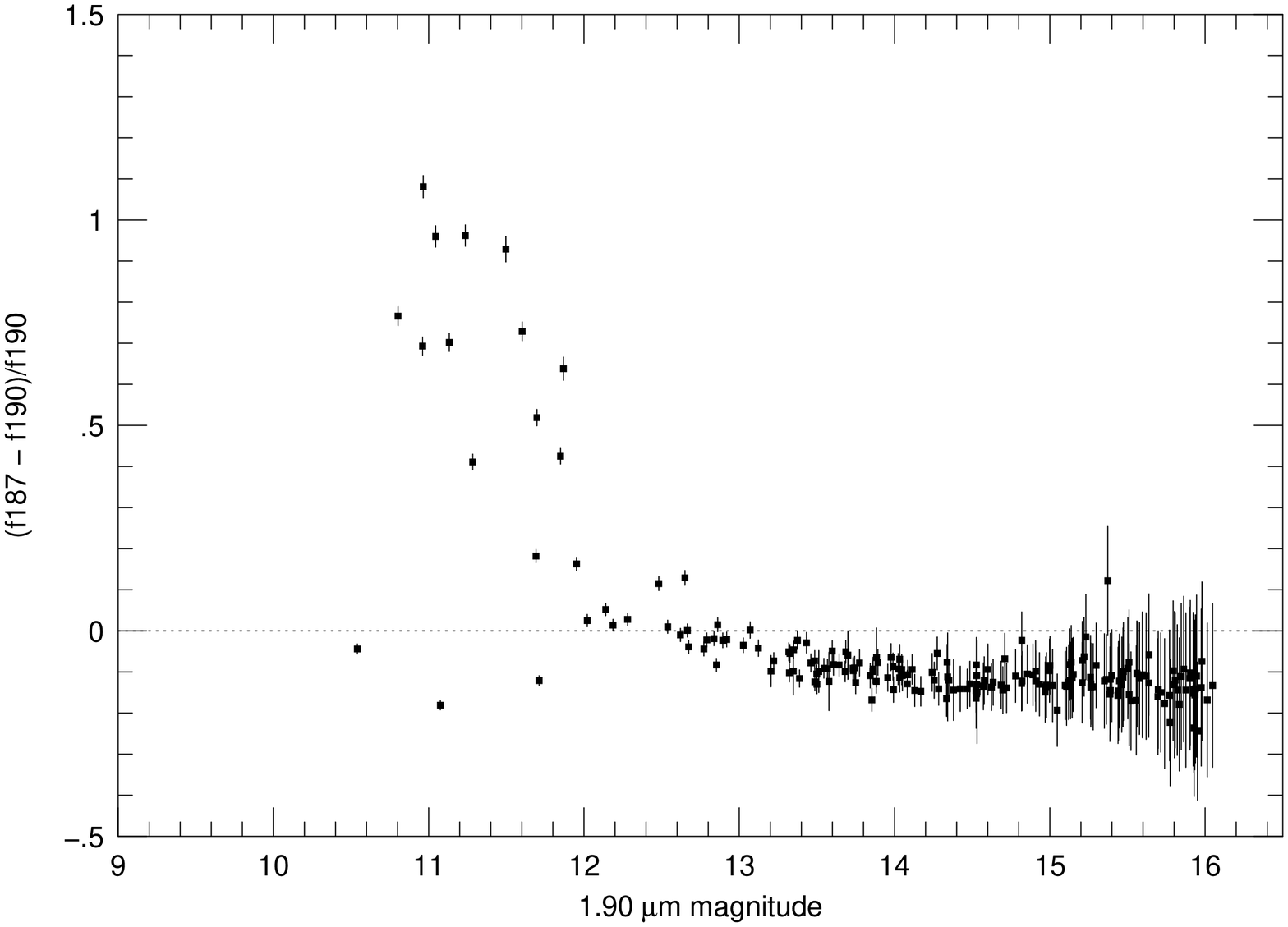} 
\figcaption[]{Same as
Figure~\ref{nb218} but for the P$\alpha$ line. This index differs from
the 2 \mic \ indices because it has no blue continuum point; see
text. The line fluxes and equivalent widths for P$\alpha$ are not
corrected for the systematic ``absorption'' evident in this plot; see
text.}
\label{nb187}
\end{figure}

\citet{ceal96} present \ion{He}{2} measurements for their stars 2 and
3. Our measurements are in good agreement with those of
\citet{ceal96}, 7.0 $\pm$ 2, and 2 $\pm$ 1 W--cm$^{-2}\times 10^{21}$
compared to 5 $\pm$ 2 and 1.2 $\pm$ 1.6 W--cm$^{-2}\times 10^{21}$.
\citet{ceal96} do not present a \ion{He}{2} measure for their star 6,
though their spectrum clearly indicates \ion{He}{2} is present. The
level appears similar to that for their star 2, and this is in
agreement with our measurements (Table~\ref{flux}). We also find a
similar high value for \ion{He}{2} for Cotera star 11. The
\citet{ceal96} spectrum is convincingly devoid of \ion{He}{2}, and
this is the same object discussed above for which the present data
showed a large ($\times$2) increase in Br$\gamma$ flux relative to
\citet{ceal96}, suggesting a real change in the atmosphere of this
star. Table~\ref{flux} and Figure~\ref{nb218} both indicate there are
a number of other stars with weak, but measurable \ion{He}{2}
emission. This is particularly evident in Figure~\ref{nb218} where the
``scatter'' in the brightest stars increases compared to that for the
intermediate brightness objects. This effect is not seen in the
\ion{He}{1} data (Figure~\ref{nb206}), suggesting that weak
\ion{He}{1} 2.06 \mic \ emission is generally absent in the
emission--line stars consistent with \citet{ceal96} and
\citet{neal95}.

Figures~\ref{brg} and \ref{paa} clearly show a new emission--line star
(22 in Table~\ref{flux}) 12.9$''$ SE of the AO guide star. The
detailed photometry (Table~\ref{flux}) confirms this. This object is
south of the region presented by \citet{ceal96}, and so was not seen
in their spectroscopic survey of the cluster. This object was noted in
broad-band images by \citet[ their object 9]{neal95}, but they did not
explicitly include it as an emission--line star nor did they present
flux measurements for it in any emission--line. A number of other,
fainter, objects present themselves as candidate emission--line
objects in Table~\ref{flux}. Number 19 has a one sigma \ion{He}{2}
line, a two sigma Br$\gamma$ line and a 20 sigma P$\alpha$ line.
Stars 8, 16, and 29 are somewhat more marginal detections, primarily
based on P$\alpha$. These are three of the four objects between 11.8
and 12.6 magnitudes in Figure~\ref{nb187} with indices between 10 and
20$\%$; the fourth is star 30. These objects are not blue, which could
otherwise lead to a false P$\alpha$ detection. Several objects, such
as numbers 20 and 31, have weak P$\alpha$ indicated , but these appear
at a level which is not significant compared to the uncertainties in
the data (see \S3.1). Numbers 19 and 22, as relatively firm
detections, are labeled ``New'' in Table~\ref{flux}; 8, 16, and 20 are
labeled as ``Candidate.''

There are several yet fainter objects indicated as emission--line
objects in Figures~\ref{nb218}, \ref{nb216}, \ref{nb206}, and
\ref{nb187}. For the ground--based data, many of these are probably
actually due to effects related to point source crowding on the
stellar photometry and follow--up spectroscopy might be performed to
confirm their nature.

\subsection{Spectral Types of the Emission--line Stars and the Age 
of the Arches Cluster}

The line strength information for the Arches emission--line stars can
be used to estimate their evolutionary state, under certain
assumptions. \citet{ceal96} asserted that the stars should be
considered as WN type based on their spectral morphology and large
measured line (velocity) widths.  An alternate possibility (mentioned
by \citet{ceal96} and \citet{coneal95}) is that one or more of the
Arches emission--line stars is a type Of supergiant.  \cite{coneal95}
presented spectra of two O4~If supergiants which have similar spectral
morphology as the late type WNs.  These two objects appear to be
similar in other respects to the late type WN stars (WNL) including
the fact that they show particularly broad emission compared to other
O~If types \citep{leep79} and may be closely related in evolutionary
status \citep{coneal95}.  It might be useful to remeasure the line
widths at higher spectral resolution. Nebular contamination on small
spatial scales and with differential velocity could cause the lines to
appear more broad. This has recently been shown to be the case in the
GC central cluster for a number of the \ion{He}{1} stars
\citep{pmmr01}.

The similarity in spectral morphology
for the Arches emission--line stars argues for a narrow range in evolutionary
state independent of the type or class which they are compared to. However,
taking the \citet{ceal96} conclusion that the Arches stars are WN
type, then the ratio of $W_{\lambda}$(2.19) to
$W_{\lambda}$(2.17) can be used to produce WN sub--types from the
tabulated values for the optically classified WN stars presented by
\citet{fmn97}. Table~\ref{equiv} gives the $W_{\lambda}$ of each
star with emission in both lines. From the ratio of these values, it
is clear that the Arches stars occupy a narrow range 
centered on the sub--type WN7. 
Only one object (21)
might have a low enough ratio to overlap with later WN8 stars;
similarly, only one object (19) has an $W_{\lambda}$ ratio between
the values of the WN6 and WN7 objects in the \citet{fmn97} catalog.

Some stars of WC sub--type have similar
$W_{\lambda}$(2.19)/$W_{\lambda}$(2.17) ratios to those for the
Arches stars \citep{fmn97}. However these WCs also typically have
either \ion{He}{1} 2.06 \mic \ emission or show strong emission in the
\ion{C}{4} triplet near 2.08 \mic, which should be detected in the
relatively broad red wing of the 2.06 \mic\ filter (\S2). The
spectroscopic survey of \citet{ceal96} also showed no evidence of
\ion{C}{4} 2.08 \mic \ or \ion{He}{1} 2.06 \mic \ emission in the
bright Arches emission--line stars. Thus, stars of WC sub--type can be
excluded from the presently identified emission--line star sample (but
see also below).

The approximate spectral types allow us to estimate the age of the
cluster.  From evolutionary models at solar metallicity an age range
between $\sim$ 2.2 and 5.6 Myr is obtained for WNL stars by
\citet{mm94} for the standard mass loss rates. Solar Fe abundances
have been measured for late--type supergiants \citep{csb00,real00} in
the GC and nearby Quintuplet clusters, though the possibility still
exists that the GC stellar population is metal rich in terms of alpha
elements.  Adopting the high mass loss models and/or models at twice
solar metallicity gives a wider range of $\sim$ 2 to 8 Myr.  There are
good indications that WN7 stars are descendants of stars with initial
masses above $\sim$ 50 -- 60 $M_\odot$ (e.g.\ \cite{sm84},
\cite{crow95}). In this case, an upper age limit of $\la$ 4--4.5 Myr
is obtained, independently of the exact metallicity and mass loss
scenario.

A similar age is also deduced if the observed emission--line stars
were O4~If stars (cf.\ above) or of the WN7~$+$abs type, which
\citet{crow95} have modeled and found intermediate between the Of and
WNL stars. \citet{bdc99}
found a WN7$+$abs star at the heart of the young stellar cluster buried
in the Galactic giant H~II region W43. 
\citet{dkeal97} concluded that two massive stars with 
Ofpe/WN6 or related WN types earlier than WN7
in the core of the R136 cluster in the LMC
were infact main--sequence objects with a maximum age of 2 Myr;
\citet{mh98} and \citet{cd98} reached the same conclusion. 
Similar type stars are present in the dense core of NGC3603 
\citep{deal95,sd99}. In R136 and NGC3603, the stars
are likely in a younger phase of evolution than the WN7 types
\citep{cd98}. The implication is thus that the Arches (and perhaps W43) is
slightly more evolved than R136 and NGC3603; a point that will be
checked in our subsequent paper when these data are used to estimate H
abundances in the Arches emission--line stars.  \citet{feal99} have
suggested that the Arches may be $\approx$ 2 $\pm$ 1 Myr old based on
near infrared photometry (consistent with the estimates discussed
here), however near infrared colors and magnitudes are insensitive to
age since even the evolved stars are on the Rayleigh--Jeans tail and
so have very small color differences.

The age range of $\sim$ 2 -- 4.5 Myr is also compatible with the
apparent absence of early type WN (WNE) and WC stars (which would
indicate older ages; cf.\ \cite{sm84}, \cite{crow95}). However, WNE
and WC stars are less luminous than WNL stars, and so it is possible
that fainter emission--line stars have been missed. The WC stars are
the faintest \citep{mm94}. \citet{bsd95} have estimated the $K-$band
magnitudes of the WC stars to be about 12.5 mag at a distance of 8
kpc, and extinction $A_K$ $=$ 3 mag, both appropriate for the Arches.
Detections of the intrinsically less luminous stars will depend not
only on the photometric completeness in each emision--line index
filter (three per line), but also on the intrinsic line strengths
which can be weak \citep{fmn97}. The present observations are similar
in angular resolution and field coverage as for the broad--band images
presented by \citet{feal99}, thus star counts for the brighter stars
which are incomplete due to crowding should also be similar between
the two data sets. \citet{feal99} find that for stars with $K$ \aple
13.5 the counts are are \apge 50$\%$ complete. Thus, if more than a
few WNE ot WC type stars with strong \ion{He}{2} 2.19 \mic \ or
\ion{He}{1} 2.06 \mic \ $+$ \ion{C}{4} \mic \ emission were present in
the cluster, a detection in either or both of Figures~\ref{nb218} and
\ref{nb206} would be expected (and depending on the object, in
Figure~\ref{nb216}, Br$\gamma$ as well), though we can not rule out
the existence of weak--lined, older stars. For example, there is an
object in Figure~\ref{nb218} at 11.5 mag with \ion{He}{2} emission
above the background. This is object no. 445 in our internal lists,
and it has $W_{\lambda}$ $=$ 21 $\pm$ 7 \AA. This object is detected
at Br$\gamma$  
2.17 \mic \ but with only 1.4 $\pm$ 4 \AA $W_{\lambda}$. The weaker
lined WNE stars presented by \citet{fmn97}, WR 116 (WN6) and WR 133
(WN4.5 $+$ O9.5 Ib), have similar \ion{He}{2} emission, but stronger
Br$\gamma$ emission.

In summary, the lack of strong lined WNE or WC detections suggests the
bulk of the cluster is represented by the age (and large number) of
the identified WNL stars, but the possibility of an older component
can not be ruled out. A detailed analysis of the properties of the
emission--line stars and comparisons with stellar models will be
presented in a forthcoming paper.

\section{Summary}

Adaptive optics and HST/NICMOS images of high angular resolution have
been presented in the emission--lines of \ion{He}{2} 2.19 \mic,
Br$\gamma$, \ion{He}{1} 2.06 \mic, and P$\alpha$. Narrow--band indices and line
fluxes have been computed for 300 to 500 stars in each line. The line
calculations will be used in a following paper to derive detailed
models for the subset of massive stars first identified by
\citet{neal95} and \citet{ceal96}. The tight sequences indicated in
narrow--band indices demonstrate that adaptive optics observations can
be used to determine accurate fluxes, even in crowded regions.

Comparison of the data presented here to previous line fluxes shows
generally good agreement in the Br$\gamma$ line, especially for the
objects farthest from the crowded core of the Arches cluster. Evidence
was presented for one object which suggests its combination of line
fluxes suggests it may have varied intrinsically between these
observations and previous ones. Other objects show larger differences,
chiefly in the nebular Br$\gamma$ line and in the more central region
of the cluster. It is suggested that these differences may be due the
difference in angular resolution between the present and past data
sets. 

Two new emission--line stars have been identified. One is a bright
object which was $\sim$ 10$''$ south of the region previously
surveyed. The other is somewhat fainter but in the more crowded region
of the central cluster, so that it was not apparent in earlier lower
angular resolution work. The narrow--band fluxes presented here
suggest these objects are similar in nature to the other massive stars
already identified in the Arches. In addition three additional
candidate stars in the cluster were identified. These are fainter
objects with somewhat weaker lines detected at lower significance.
Significant differences in the spectroscopic character of Cotera star
11 indicate that real changes may have occured between the earlier
observations (1992/1993) of \citet{ceal96} and those presented here.

Detailed spectral types are presented for the Arches emission--line
stars with 2.19 \mic \ and 2.17 \mic \ emission under the assumption
that they are Wolf--Rayet stars. Irrespective of the precise spectral
type, the emission--line character of the bright cluster stars is
remarkably similar, hence the evolutionary spread is quite narrow.
>From the approximate spectral types of the identified emission--line
stars and comparisons with evolutionary models we infer a cluster age
between $\sim$ 2 and 4.5 Myr.

The authors would like to thank the staff of the CFHT for their
excellent support while at the telescope and Fran\c{c}ois M\'enard for
assistance with filter transmission curves.  The authors also
acknowledge D. Figer for taking the original P$\alpha$ data as part of
his HST/NICMOS Arches program. Finally, the authors thank A. Cotera
for useful discussions about the calibration of the HST images.

%REFERENCES

\newpage

%FIGURES

\begin{figure}
\figurenum{1$a$} 
\plotone{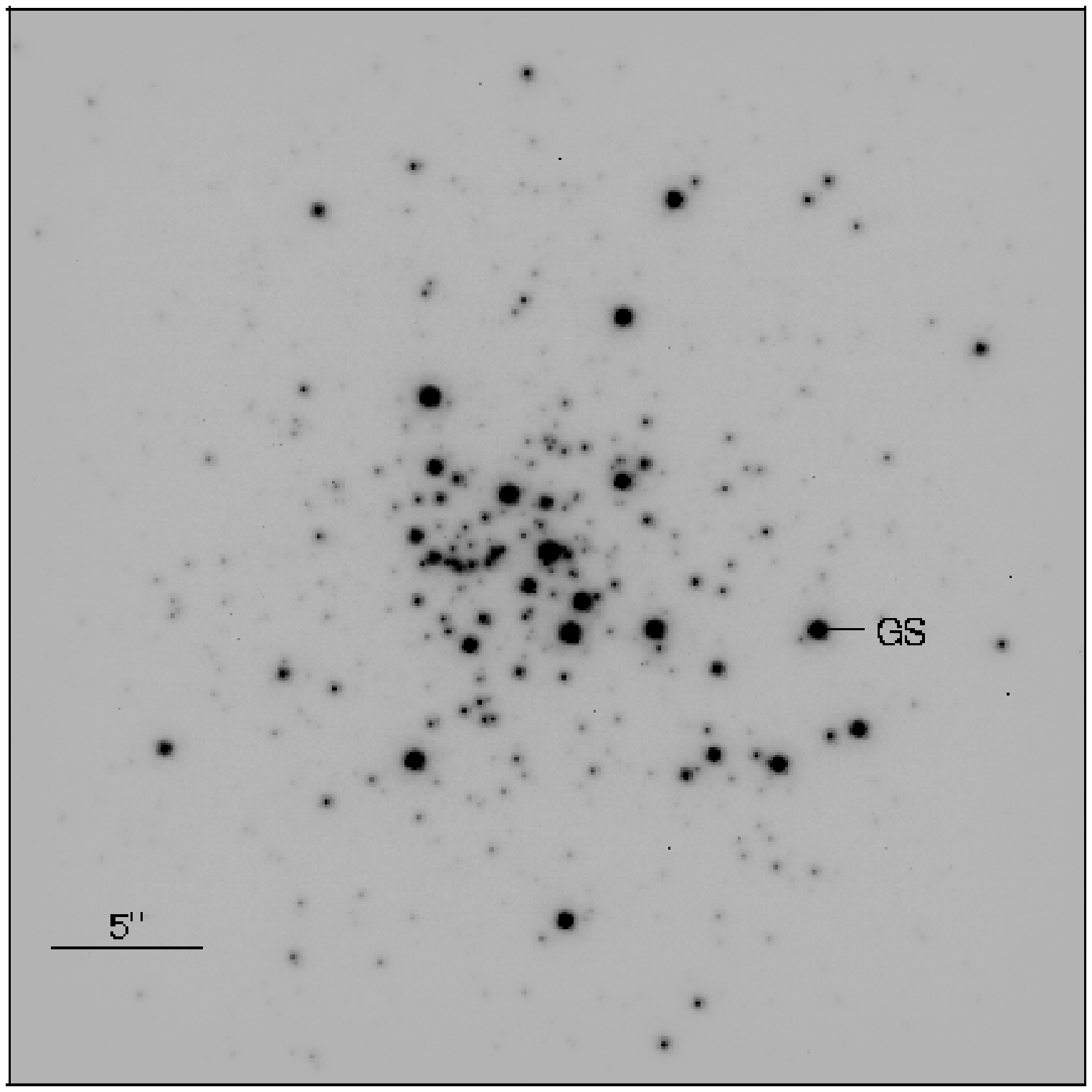} 
\figcaption[]{
\ion{He}{2} 2.19 \mic \ CFHT adaptive optics bonnette image 
of the Arches cluster. 
North is up, East to the left, and the scale is 0.035$''$ pix$^{-1}$ in 
this $\sim$ 36$''$ $\times$ 36$''$ image. The adaptive optics guide star is
labeled ``GS.''}
\label{a218}
\end{figure}

\begin{figure}
\figurenum{1$b$}
\plotone{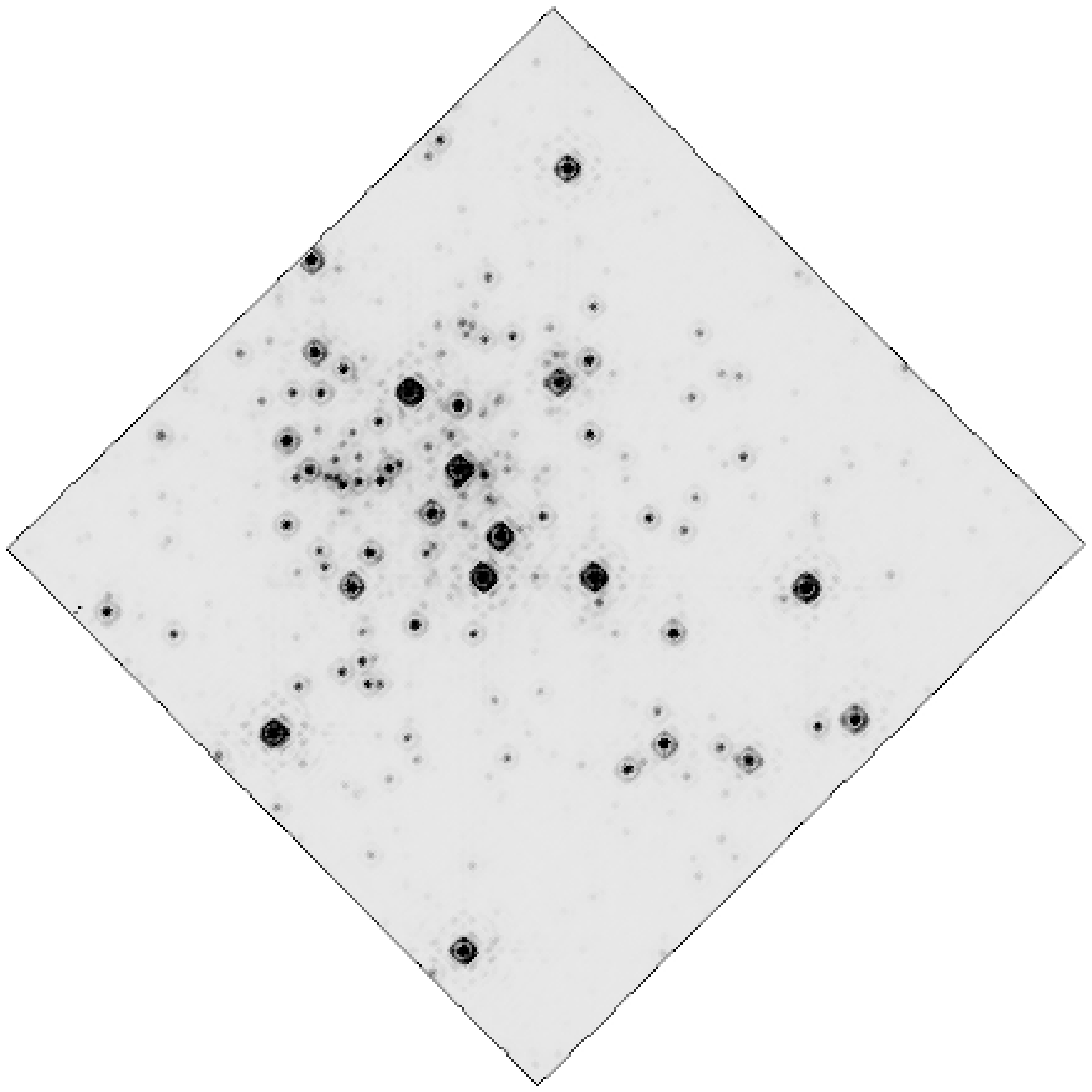}
\figcaption[]{F187 (P$\alpha$) HST/NICMOS image. North up and East to 
the left.
This image has been rotated by -44.7 degrees counter--clockwise from the 
original orientation.
The scale is 0.076$''$ pix$^{-1}$ in this 19$''\times19''$ image.}
\label{f187}
\end{figure}

\begin{figure}
\figurenum{8} 
\plotone{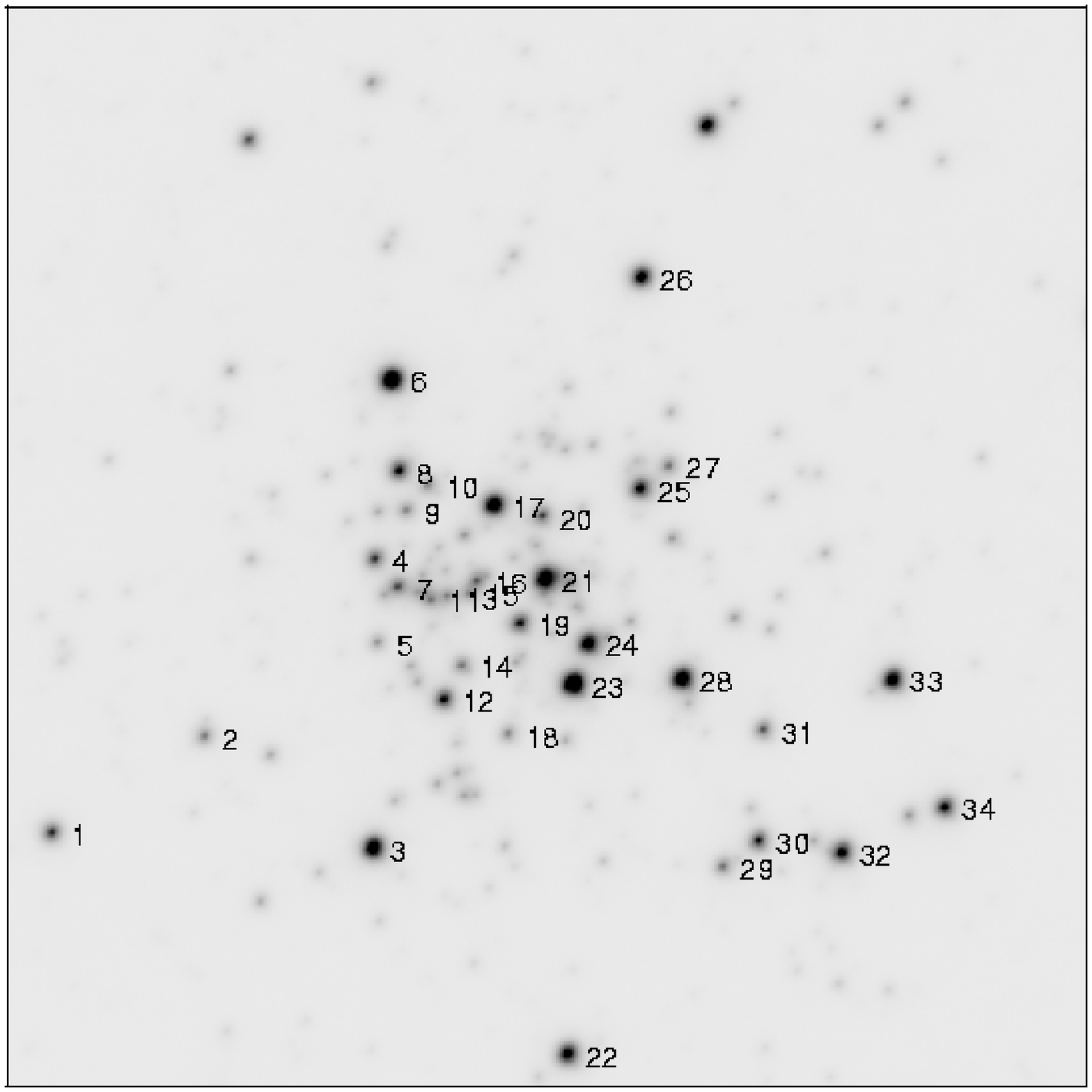} 

\figcaption[]{Positions of stars corresponding to the IDs in
Tables~\ref{pos}, \ref{mags}, and \ref{flux} overlaid on the 2.26 \mic
\ image. The region shown is smaller (28$''\times28''$) 
than in Figure~\ref{a218}, but
the orientation is the same. Corresponding IDs for \citet{neal95} and
\citet{ceal96} are given in Table~\ref{flux}.}

\label{a226}
\end{figure}

\newpage

%TABLES
%WRTAB

\pagestyle{empty}
\begin{deluxetable}{lcrrrrrrrrrr}
\rotate
\tablecaption{Wolf--Rayet Star Observed Line Strengths (\AA) \label{wr}} 
\tablewidth{0pt}
\tablehead{ 
\colhead{Star} & 
\colhead{Sp Type} &
\colhead{\ion{He}{1} 2.06 \mic} &
\colhead{Lit\tablenotemark{b} } &
\colhead{$\Delta$\tablenotemark{c}} &
\colhead{Br$\gamma$ 2.17 \mic  } & 
\colhead{Lit \tablenotemark{b}} &
\colhead{$\Delta$\tablenotemark{c}} &
\colhead{\ion{He}{2} 2.19 \mic } &
\colhead{Lit \tablenotemark{b}} &
\colhead{$\Delta$} &
\colhead{\ion{He}{2}/Br$\gamma$}  
}
\startdata
WR122 &Ofpe/WN9&623\tablenotemark{a}&680&$-$0.08&25&22& 0.14 &\nodata&\nodata&\nodata \\
WR123&WN8 &48\tablenotemark{a} &    77& $-$0.38&  65&74&$-$0.12 & 26 &19& 0.37 &0.40\\
WR128&WN4 & \nodata& \nodata&\nodata&  43  & 44  &$-$0.02 & 63  & 76 &$-$0.17&1.47\\
WR131&WN7+abs & \nodata& \nodata&\nodata&47&41&0.15&220& 18&\nodata\tablenotemark{d} &4.68\\
WR134&WN6 & \nodata& \nodata&\nodata&  75 & 83 &$-$0.10 & 129 & 137 &$-$0.06 &1.72\\
WR138&WN5+O9 & \nodata& \nodata& \nodata& 49 & 35 &   0.40  & 65  & 54 & 0.20 & 1.33\\
mean  &  & & & & &  0.07 &  &         &0.09 \\
stddev&  & & & & &  0.08 &  &         &0.12 \\
\enddata 

\tablenotetext{a}{Continuum is based on 2.14 \mic \ point only; see
text}.

\tablenotetext{b}{Values for equivalent widths from spectroscopic
measurements are from \citet{bds95}, \citet{cs96}, and \citet{fmn97}.}

\tablenotetext{c}{Difference in equivalent width as a percentage of
the literature value}

\tablenotetext{d}{The narrow--band \ion{He}{2} line for WR131 is an
order of magnitude stronger than the literature value \citep{fmn97};
see discussion in \S2.1.3}

\end{deluxetable}

\newpage

\pagestyle{empty}
\begin{deluxetable}{lccc}
\tablecaption{Arches Emission--Line Star Positions\label{pos}}
\tablewidth{0pt}
\tablehead{
\colhead{Star ID} &
\colhead{RA ($''$)\tablenotemark{a}} &
\colhead{DEC ($''$)\tablenotemark{a}} &
\colhead{Radius ($''$)\tablenotemark{a}} }
\startdata
1        &21.8   &$-$4.0   &22.2 \\
2        &17.9   &$-$1.5   &17.9  \\
3        &13.5   &$-$4.4   &14.2  \\
4        &13.4   &3.1    &13.8  \\
5        &13.4   &1.0    &13.4  \\
6        &13.0   &7.8    &15.1  \\
7        &12.8   &2.4    &13.1  \\
8        &12.8   &5.4    &13.9  \\
9        &12.6   &4.4    &13.4  \\
10       &12.1   &5.0    &13.1  \\
11       &12.0   &2.1    &12.2  \\
12       &11.6   &$-$0.5   &11.6  \\
13       &11.6   &2.2    &11.8  \\
14       &11.2   &0.4    &11.2  \\
15       &11.0   &2.2    &11.2  \\
16       &10.8   &2.6    &11.1  \\
17       &10.3   &4.5    &11.3  \\
18       &10.0   &$-$1.4   &10.1  \\
19       &9.7    &1.5    &9.8  \\
20       &9.1    &4.3    &10.0  \\
21       &9.0    &2.6    &9.4  \\
22       &8.4    &$-$9.7   &12.9  \\
23       &8.3    &$-$0.1   &8.3  \\
24       &7.9    &0.9    &8.0  \\
25       &6.5    &5.0    &8.2  \\
26       &6.5    &10.5   &12.3  \\
27       &5.8    &5.6    &8.0  \\
28       &5.5    &0.0    &5.5  \\
29       &4.4    &$-$4.9   &6.6  \\
30       &3.5    &$-$4.2   &5.4  \\
31       &3.4    &$-$1.3   &3.6  \\
32       &1.3    &$-$4.5   &4.7  \\
33       &0.0    &0.0    &0.0  \\
34       &$-$1.4   &$-$3.3   &3.6  \\
\enddata

\tablenotetext{a}{Offset from the adaptive optics guide star, Star ID 33;
see also Figure~\ref{a218}.}

\end{deluxetable}

\newpage

%MAGS
\pagestyle{empty}
\begin{deluxetable}{lcccccccc}
\tablecaption{Arches Emission--Line Star Narrow--band Magnitudes\label{mags}}
\tablewidth{0pt}
\rotate
\tablehead{
\colhead{Star ID} &
\colhead{2.26 \mic} &
\colhead{2.19 \mic} &
\colhead{2.17 \mic} &
\colhead{2.14 \mic} &
\colhead{2.06 \mic} &
\colhead{2.03 \mic} &
\colhead{1.90 \mic} &
\colhead{1.87 \mic} }
\startdata
     1&   11.08$\pm$    0.01&   11.10$\pm$    0.02&   11.03$\pm$    0.01&   11.11$\pm$    0.03&   11.29$\pm$    0.03&   11.34$\pm$    0.02& \nodata&
 nodata\\
     2&   11.78$\pm$    0.01&   11.79$\pm$    0.02&   11.80$\pm$    0.01&   11.76$\pm$    0.03&   11.88$\pm$    0.03&   11.91$\pm$    0.02&   12.67$
\pm$    0.01&   12.69$\pm$    0.01\\
     3&   10.13$\pm$    0.01&   10.09$\pm$    0.02&    9.94$\pm$    0.01&   10.10$\pm$    0.03&   10.29$\pm$    0.03&   10.32$\pm$    0.02&   11.05$
\pm$    0.01&   10.34$\pm$    0.01\\
     4&   11.15$\pm$    0.01&   11.19$\pm$    0.02&   11.17$\pm$    0.01&   11.12$\pm$    0.03&   11.27$\pm$    0.03&   11.27$\pm$    0.02&   12.02$
\pm$    0.01&   12.02$\pm$    0.01\\
     5&   12.10$\pm$    0.01&   12.11$\pm$    0.02&   12.11$\pm$    0.01&   12.05$\pm$    0.03&   12.22$\pm$    0.03&   12.21$\pm$    0.02&   12.92$
\pm$    0.01&   12.97$\pm$    0.01\\
     6&    9.80$\pm$    0.01&    9.86$\pm$    0.02&    9.87$\pm$    0.01&    9.88$\pm$    0.03&   10.20$\pm$    0.03&   10.28$\pm$    0.02&   11.07$
\pm$    0.01&   11.31$\pm$    0.01\\
     7&   11.35$\pm$    0.01&   11.37$\pm$    0.02&   11.38$\pm$    0.01&   11.31$\pm$    0.03&   11.47$\pm$    0.03&   11.49$\pm$    0.02&   12.19$
\pm$    0.01&   12.19$\pm$    0.01\\
     8&   10.82$\pm$    0.01&   10.84$\pm$    0.02&   10.82$\pm$    0.02&   10.78$\pm$    0.03&   10.93$\pm$    0.03&   10.92$\pm$    0.02&   11.69$
\pm$    0.01&   11.53$\pm$    0.01\\
     9&   11.87$\pm$    0.01&   11.90$\pm$    0.02&   11.91$\pm$    0.01&   11.83$\pm$    0.03&   11.97$\pm$    0.03&   11.98$\pm$    0.02&   12.67$
\pm$    0.01&   12.74$\pm$    0.01\\
    10&   11.94$\pm$    0.02&   11.99$\pm$    0.02&   11.99$\pm$    0.03&   11.92$\pm$    0.03&   12.04$\pm$    0.03&   12.06$\pm$    0.02&   12.79$
\pm$    0.01&   12.84$\pm$    0.01\\
    11&   12.00$\pm$    0.01&   12.02$\pm$    0.02&   12.03$\pm$    0.01&   11.96$\pm$    0.03&   12.08$\pm$    0.03&   12.09$\pm$    0.02&   12.84$
\pm$    0.01&   12.88$\pm$    0.01\\
    12&   10.80$\pm$    0.01&   10.79$\pm$    0.02&   10.71$\pm$    0.01&   10.76$\pm$    0.03&   10.94$\pm$    0.03&   10.96$\pm$    0.02&   11.70$
\pm$    0.01&   11.27$\pm$    0.01\\
    13&   12.02$\pm$    0.01&   12.03$\pm$    0.02&   12.05$\pm$    0.01&   11.98$\pm$    0.03&   12.11$\pm$    0.03&   12.14$\pm$    0.02&   12.86$
\pm$    0.01&   12.87$\pm$    0.01\\
    14&   11.80$\pm$    0.01&   11.80$\pm$    0.02&   11.85$\pm$    0.01&   11.75$\pm$    0.03&   11.90$\pm$    0.03&   11.91$\pm$    0.02&   12.62$
\pm$    0.01&   12.65$\pm$    0.01\\
    15&   12.09$\pm$    0.01&   12.09$\pm$    0.02&   12.15$\pm$    0.01&   12.05$\pm$    0.03&   12.17$\pm$    0.03&   12.18$\pm$    0.02&   12.89$
\pm$    0.01&   12.94$\pm$    0.01\\
    16&   11.64$\pm$    0.01&   11.65$\pm$    0.02&   11.67$\pm$    0.01&   11.60$\pm$    0.03&   11.75$\pm$    0.03&   11.77$\pm$    0.02&   12.48$
\pm$    0.01&   12.39$\pm$    0.01\\
    17&   10.10$\pm$    0.01&   10.04$\pm$    0.02&    9.92$\pm$    0.01&   10.06$\pm$    0.03&   10.27$\pm$    0.03&   10.27$\pm$    0.02&   10.97$
\pm$    0.01&   10.19$\pm$    0.01\\
    18&   11.92$\pm$    0.01&   11.92$\pm$    0.02&   11.96$\pm$    0.01&   11.89$\pm$    0.03&   12.06$\pm$    0.03&   12.07$\pm$    0.02&   12.77$
\pm$    0.01&   12.84$\pm$    0.01\\
    19&   10.98$\pm$    0.01&   10.96$\pm$    0.02&   10.94$\pm$    0.01&   10.93$\pm$    0.03&   11.11$\pm$    0.03&   11.13$\pm$    0.02&   11.85$
\pm$    0.01&   11.49$\pm$    0.01\\
    20&   11.34$\pm$    0.01&   11.33$\pm$    0.02&   11.36$\pm$    0.01&   11.28$\pm$    0.03&   11.43$\pm$    0.03&   11.43$\pm$    0.02&   12.14$
\pm$    0.01&   12.11$\pm$    0.01\\
    21&   10.05$\pm$    0.01&   10.04$\pm$    0.02&    9.94$\pm$    0.01&   10.02$\pm$    0.03&   10.22$\pm$    0.03&   10.22$\pm$    0.02&   10.96$
\pm$    0.01&   10.41$\pm$    0.01\\
    22&   10.50$\pm$    0.01&   10.44$\pm$    0.02&   10.32$\pm$    0.01&   10.47$\pm$    0.03&   10.64$\pm$    0.03&   10.71$\pm$    0.02&   11.50$
\pm$    0.01&   10.81$\pm$    0.01\\
    23&    9.80$\pm$    0.01&    9.76$\pm$    0.02&    9.67$\pm$    0.01&    9.76$\pm$    0.03&    9.98$\pm$    0.03&    9.98$\pm$    0.02&   10.80$
\pm$    0.01&   10.21$\pm$    0.01\\
    24&   10.26$\pm$    0.01&   10.22$\pm$    0.02&   10.10$\pm$    0.01&   10.23$\pm$    0.03&   10.45$\pm$    0.03&   10.45$\pm$    0.02&   11.24$
\pm$    0.01&   10.53$\pm$    0.01\\
    25&   10.66$\pm$    0.01&   10.64$\pm$    0.02&   10.56$\pm$    0.01&   10.63$\pm$    0.03&   10.82$\pm$    0.03&   10.82$\pm$    0.02&   11.60$
\pm$    0.01&   11.03$\pm$    0.01\\
    26&   10.43$\pm$    0.01&   10.41$\pm$    0.02&   10.36$\pm$    0.01&   10.39$\pm$    0.03&   10.55$\pm$    0.03&   10.56$\pm$    0.02&   11.28$
\pm$    0.01&   10.93$\pm$    0.01\\
    27&   11.68$\pm$    0.01&   11.72$\pm$    0.02&   11.70$\pm$    0.03&   11.64$\pm$    0.03&   11.81$\pm$    0.03&   11.82$\pm$    0.03&   12.54$
\pm$    0.01&   12.55$\pm$    0.01\\
    28&   10.13$\pm$    0.01&   10.12$\pm$    0.02&   10.05$\pm$    0.01&   10.11$\pm$    0.03&   10.33$\pm$    0.03&   10.34$\pm$    0.02&   11.13$
\pm$    0.01&   10.58$\pm$    0.01\\
    29&   11.66$\pm$    0.01&   11.68$\pm$    0.02&   11.67$\pm$    0.01&   11.64$\pm$    0.03&   11.83$\pm$    0.03&   11.85$\pm$    0.02&   12.65$
\pm$    0.01&   12.54$\pm$    0.01\\
    30&   10.98$\pm$    0.01&   11.00$\pm$    0.02&   11.00$\pm$    0.01&   10.97$\pm$    0.03&   11.16$\pm$    0.03&   11.18$\pm$    0.02&   11.95$
\pm$    0.01&   11.81$\pm$    0.01\\
    31&   11.37$\pm$    0.01&   11.40$\pm$    0.02&   11.43$\pm$    0.01&   11.35$\pm$    0.03&   11.54$\pm$    0.03&   11.55$\pm$    0.02&   12.28$
\pm$    0.01&   12.27$\pm$    0.01\\
    32&   10.63$\pm$    0.01&   10.65$\pm$    0.02&   10.69$\pm$    0.01&   10.65$\pm$    0.03&   10.91$\pm$    0.03&   10.96$\pm$    0.02&   11.71$
\pm$    0.01&   11.87$\pm$    0.01\\
    33&   10.35$\pm$    0.01&   10.24$\pm$    0.02&   10.25$\pm$    0.01&   10.10$\pm$    0.03&   10.10$\pm$    0.03&   10.07$\pm$    0.02&   10.54$
\pm$    0.01&   10.61$\pm$    0.01\\
    34&   10.74$\pm$    0.01&   10.76$\pm$    0.02&   10.69$\pm$    0.01&   10.78$\pm$    0.03&   11.03$\pm$    0.03&   11.06$\pm$    0.02&   11.87$
\pm$    0.02&   11.35$\pm$    0.01\\
\enddata
\end{deluxetable}

\newpage

%FLUXES
\pagestyle{empty}
\begin{deluxetable}{llrrrr}
\tablecaption{Arches Emission--Line Star Line Fluxes (W--cm$^{-2}$ 
$\times$ $10^{21}$)\label{flux}}
\tablewidth{0pt}
\tablehead{
\colhead{Star ID} &
\colhead{Other ID\tablenotemark{a}} &
\colhead{\ion{He}{2} 2.19 \mic\tablenotemark{b}} &
\colhead{Br$\gamma$ \tablenotemark{b}} &
\colhead{\ion{He}{1} 2.06 \mic\tablenotemark{b}} &
\colhead{P$\alpha$} }
\startdata
     1&10&    1.43$\pm$    0.89&    3.12$\pm$    0.63&    0.05$\pm$    1.23&    \nodata         \\
     2& &   0.31$\pm$    0.47&    0.00$\pm$    0.31&    0.23$\pm$    0.68&    0.00$\pm$    0.22\\
     3& 11, N14&   5.69$\pm$    2.25&   15.51$\pm$    1.72&   $-$0.75$\pm$    3.12&   46.79$\pm$    1.46\\
     4& 4&  $-$1.12$\pm$    0.82&    0.00$\pm$    0.55&   $-$0.31$\pm$    1.22&    0.60$\pm$    0.20\\
     5&  & $-$0.12$\pm$    0.35&   $-$0.12$\pm$    0.23&   $-$0.43$\pm$    0.52&   $-$0.17$\pm$    0.17\\
     6& A, N13&   3.17$\pm$    2.78&    2.67$\pm$    1.84&   $-$1.79$\pm$    3.82&   $-$8.61$\pm$    0.48\\
     7& &  $-$0.59$\pm$    0.69&   $-$0.47$\pm$    0.46&   $-$0.25$\pm$    1.02&    0.17$\pm$    0.17\\
     8& B, N12, Candidate&  $-$0.38$\pm$    1.13&    0.38$\pm$    0.76&   $-$1.42$\pm$    1.67&    4.87$\pm$    0.54\\
     9& &  $-$0.36$\pm$    0.42&   $-$0.29$\pm$    0.28&   $-$0.16$\pm$    0.63&   $-$0.44$\pm$    0.22\\
    10&  & $-$0.33$\pm$    0.39&   $-$0.13$\pm$    0.39&    0.02$\pm$    0.58&   $-$0.20$\pm$    0.20\\
    11& &  $-$0.13$\pm$    0.38&   $-$0.26$\pm$    0.25&   $-$0.15$\pm$    0.56&   $-$0.19$\pm$    0.19\\
    12&  7&  1.34$\pm$    1.18&    3.85$\pm$    0.85&   $-$0.41$\pm$    1.70&   13.93$\pm$    0.54\\
    13&&    0.06$\pm$    0.38&   $-$0.12$\pm$    0.25&    0.18$\pm$    0.55&    0.09$\pm$    0.18\\
    14& &   0.08$\pm$    0.47&   $-$0.46$\pm$    0.30&   $-$0.17$\pm$    0.68&   $-$0.11$\pm$    0.23\\
    15&  &  0.06$\pm$    0.36&   $-$0.47$\pm$    0.22&   $-$0.13$\pm$    0.52&   $-$0.18$\pm$    0.18\\
    16&Candidate & 0.09$\pm$    0.53&   $-$0.18$\pm$    0.35&    0.03$\pm$    0.78&    1.57$\pm$    0.26\\
    17&  2, N11&  7.01$\pm$    2.35&   13.37$\pm$    1.75&   $-$3.57$\pm$    3.23&   56.67$\pm$    1.57\\
    18&  &  0.27$\pm$    0.42&   $-$0.27$\pm$    0.27&   $-$0.32$\pm$    0.60&   $-$0.40$\pm$    0.20\\
    19& New &  1.15$\pm$    1.01&    1.32$\pm$    0.68&   $-$0.35$\pm$    1.45&   10.03$\pm$    0.47\\
    20& &   0.12$\pm$    0.72&   $-$0.47$\pm$    0.47&   $-$0.26$\pm$    1.05&    0.89$\pm$    0.18\\
    21& 5, N10&   2.68$\pm$    2.35&    9.22$\pm$    1.72&   $-$2.74$\pm$    3.36&   36.54$\pm$    1.06\\
    22& New&   4.85$\pm$    1.63&   10.10$\pm$    1.21&    1.33$\pm$    2.22&   29.95$\pm$    0.97\\
    23&8, N8 &   6.33$\pm$    3.05&   13.55$\pm$    2.21&   $-$3.41$\pm$    4.26&   47.25$\pm$    1.23\\
    24&6, N7&    5.05$\pm$    1.99&   10.79$\pm$    1.48&   $-$3.02$\pm$    2.77&   39.28$\pm$    1.23\\
    25&3, N6&    2.18$\pm$    1.35&    5.21$\pm$    0.97&   $-$1.57$\pm$    1.91&   21.44$\pm$    0.59\\
    26&1, N5&    1.91$\pm$    1.67&    4.33$\pm$    1.17&   $-$1.30$\pm$    2.39&   16.17$\pm$    0.79\\
    27&&   $-$0.69$\pm$    0.50&    0.00$\pm$    0.51&   $-$0.19$\pm$    0.75&    0.12$\pm$    0.25\\
    28&9, N4&    2.49$\pm$    2.19&    7.07$\pm$    1.55&   $-$2.47$\pm$    3.09&   31.70$\pm$    0.91\\
    29&Candidate &   0.09$\pm$    0.52&    0.17$\pm$    0.35&   $-$0.40$\pm$    0.75&    1.45$\pm$    0.22\\
    30&12&    0.16$\pm$    0.97&    0.32$\pm$    0.65&   $-$0.74$\pm$    1.40&    3.40$\pm$    0.43\\
    31&E&   $-$0.23$\pm$    0.67&   $-$0.45$\pm$    0.44&   $-$0.52$\pm$    0.99&    0.47$\pm$    0.16\\
    32&  G, N3 & 1.53$\pm$    1.34&    0.00$\pm$    0.86&   $-$0.93$\pm$    1.88&   $-$3.18$\pm$    0.27\\
    33& F, N2, GS\tablenotemark{c}&  $-$2.68$\pm$    1.96&   $-$2.69$\pm$    1.29&    0.15$\pm$    3.12&   $-$3.12$\pm$    0.78\\
    34& 13, N1&   2.52$\pm$    1.21&    4.27$\pm$    0.86&   $-$1.30$\pm$    1.67&   14.66$\pm$    0.69\\
\enddata

\tablenotetext{a}{Emission--line star ID from \citet{ceal96},
\citet{neal95} (``N''), or other ID from this paper.}

\tablenotetext{b}{Flux has been corrected for intrinsic absorption in
O and B--type stars; see \S3.1.}

\tablenotetext{c}{Adaptive optics guide star; see Figure~\ref{a218}.}

\end{deluxetable}

\pagestyle{empty}
\begin{deluxetable}{lrrrrrr}
\tablecaption{Arches Emission--Line Star Line Equivalent Widths 
(\AA)\label{equiv}}
\tablewidth{0pt}
\tablehead{
\colhead{Star ID} &
\colhead{\ion{He}{2} 2.19 \mic\tablenotemark{a}} &
\colhead{Br$\gamma$ \tablenotemark{a}} &
\colhead{\ion{He}{1} 2.06 \mic\tablenotemark{a}} &
\colhead{P$\alpha$\tablenotemark{a}} &
\colhead{\ion{He}{2} 2.19 \mic/Br$\gamma$}   &
\colhead{SpTyp \tablenotemark{b}}
}
\startdata
     1&   10.00$\pm$    6.00&   22.00$\pm$    4.00&    0.40$\pm$    8.40&    0.00$\pm$    0.00&    0.45&WN7   \\
     2&    4.00$\pm$    6.00&    0.00$\pm$    4.00&    3.20$\pm$    8.40&    0.00$\pm$    3.70&    \nodata&\nodata\\
     3&   16.00$\pm$    6.00&   44.00$\pm$    4.00&   $-$2.40$\pm$    8.40&  179.50$\pm$    5.60&    0.36&WN7   \\
     4&   $-$8.00$\pm$    6.00&    0.00$\pm$    4.00&   $-$2.40$\pm$    8.40&    5.60$\pm$    1.90&    \nodata&\nodata\\
     5&   $-$2.00$\pm$    6.00&   $-$2.00$\pm$    4.00&   $-$8.00$\pm$    8.40&   $-$3.70$\pm$    3.70&   \nodata &\nodata\\
     6&    7.00$\pm$    6.00&    6.00$\pm$    4.00&   $-$5.20$\pm$    8.40&  $-$33.70$\pm$    1.90&    \nodata&\nodata\\
     7&   $-$5.00$\pm$    6.00&   $-$4.00$\pm$    4.00&   $-$2.40$\pm$    8.40&    1.90$\pm$    1.90&    \nodata&\nodata\\
     8&   $-$2.00$\pm$    6.00&    2.00$\pm$    4.00&   $-$8.00$\pm$    8.40&   33.70$\pm$    3.70&    &\nodata\\
     9&   $-$5.00$\pm$    6.00&   $-$4.00$\pm$    4.00&   $-$2.40$\pm$    8.40&   $-$7.50$\pm$    3.70&    \nodata&\nodata\\
    10&   $-$5.00$\pm$    6.00&   $-$2.00$\pm$    6.00&    0.40$\pm$    8.40&   $-$3.70$\pm$    3.70&   \nodata &\nodata\\
    11&   $-$2.00$\pm$    6.00&   $-$4.00$\pm$    4.00&   $-$2.40$\pm$    8
.40&   $-$3.70$\pm$    3.70&    \nodata&\nodata\\
    12&    7.00$\pm$    6.00&   20.00$\pm$    4.00&   $-$2.40$\pm$    8.40&   97.20$\pm$    3.70&    0.35&WN7   \\
    13&    1.00$\pm$    6.00&   $-$2.00$\pm$    4.00&    3.20$\pm$    8.40&    1.90$\pm$    3.70&    \nodata&\nodata\\
    14&    1.00$\pm$    6.00&   $-$6.00$\pm$    4.00&   $-$2.40$\pm$    8.40&   $-$1.90$\pm$    3.70&   \nodata &\nodata\\
    15&    1.00$\pm$    6.00&   $-$8.00$\pm$    4.00&   $-$2.40$\pm$    8.40&   $-$3.70$\pm$    3.70&    \nodata&\nodata\\
    16&    1.00$\pm$    6.00&   $-$2.00$\pm$    4.00&    0.40$\pm$    8.40&   22.40$\pm$    3.70&    \nodata&\nodata\\
    17&   19.00$\pm$    6.00&   36.00$\pm$    4.00&  $-$10.80$\pm$    8.40&  202.00$\pm$    5.60&    0.53&WN7   \\
    18&    4.00$\pm$    6.00&   $-$4.00$\pm$    4.00&   $-$5.20$\pm$    8.40&   $-$7.50$\pm$    3.70&    \nodata&\nodata\\
    19&    7.00$\pm$    6.00&    8.00$\pm$    4.00&   $-$2.40$\pm$    8.40&   80.40$\pm$    3.70&    0.88&WN6-7   \\
    20&    1.00$\pm$    6.00&   $-$4.00$\pm$    4.00&   $-$2.40$\pm$    8.40&    9.40$\pm$    1.90&    \nodata&\nodata\\
    21&    7.00$\pm$    6.00&   24.00$\pm$    4.00&   $-$8.00$\pm$    8.40&  129.00$\pm$    3.70&    0.29&WN7-8   \\
    22&   19.00$\pm$    6.00&   40.00$\pm$    4.00&    6.00$\pm$    8.40&  173.90$\pm$    5.60&    0.48&WN7   \\
    23&   13.00$\pm$    6.00&   28.00$\pm$    4.00&   $-$8.00$\pm$    8.40&  144.00$\pm$    3.70&    0.46&WN7   \\
    24&   16.00$\pm$    6.00&   34.00$\pm$    4.00&  $-$10.80$\pm$    8.40&  179.50$\pm$    5.60&    0.47&WN7   \\
    25&   10.00$\pm$    6.00&   24.00$\pm$    4.00&   $-$8.00$\pm$    8.40&  136.50$\pm$    3.70&    0.42&WN7   \\
    26&    7.00$\pm$    6.00&   16.00$\pm$    4.00&   $-$5.20$\pm$    8.40&   76.70$\pm$    3.70&    0.44&WN7   \\
    27&   $-$8.00$\pm$    6.00&    0.00$\pm$    6.00&   $-$2.40$\pm$    8.40&    1.90$\pm$    3.70&    \nodata&\nodata\\
    28&    7.00$\pm$    6.00&   20.00$\pm$    4.00&   $-$8.00$\pm$    8.40&  130.90$\pm$    3.70&    0.35&WN7   \\
    29&    1.00$\pm$    6.00&    2.00$\pm$    4.00&   $-$5.20$\pm$    8.40&   24.30$\pm$    3.70&    0.50&WN7   \\
    30&    1.00$\pm$    6.00&    2.00$\pm$    4.00&   $-$5.20$\pm$    8.40&   29.90$\pm$    3.70&    0.50&WN7   \\
    31&   $-$2.00$\pm$    6.00&   $-$4.00$\pm$    4.00&   $-$5.20$\pm$    8.40&    5.60$\pm$    1.90&    \nodata&\nodata\\
    32&    7.00$\pm$    6.00&    0.00$\pm$    4.00&   $-$5.20$\pm$    8.40&  $-$22.40$\pm$    1.90&    \nodata&\nodata\\
    33&   $-$8.00$\pm$    6.00&   $-$8.00$\pm$    4.00&    0.40$\pm$    8.40&   $-$7.50$\pm$    1.90&    \nodata&\nodata\\
    34&   13.00$\pm$    6.00&   22.00$\pm$    4.00&   $-$8.00$\pm$    8.40&  119.70$\pm$    5.60&    0.59&WN7   \\
\enddata

\tablenotetext{a}{Equivalent width is calculated as the narrow--band
index multiplied by the filter bandpass; see \S2.1.3}

\tablenotetext{b}{SpTyp from \citet{fmn97} assuming the stars are
Wolf--Rayet type \citep{ceal96} and not Of supergiants; see text}.
\end{deluxetable}

\end{document}